\documentclass[
prl,aps,
reprint,
a4paper,
superscriptaddress,
longbibliography,
floatfix
]{revtex4-1}
\usepackage{amsmath,amsthm,amsfonts}
\usepackage{amssymb}
\usepackage{braket}
\usepackage{epsfig}
\usepackage{graphicx}
\usepackage{multirow}
\usepackage{times}
\usepackage{units}
\usepackage{gensymb}
\usepackage[colorlinks=true,linkcolor=blue,citecolor=magenta,urlcolor=blue]{hyperref}
\usepackage{array}

\newcolumntype{L}[1]{>{\raggedright\let\newline\\\arraybackslash\hspace{0pt}}m{#1}}

\begin{document}


\title{Observation of Stronger-than-Binary Correlations with Entangled Photonic 
Qutrits}



\author{Xiao-Min~Hu}
\affiliation{CAS Key Laboratory of Quantum Information, University of Science 
and Technology of China, Hefei, 230026, People's Republic of China}
\affiliation{Synergetic Innovation Center of Quantum Information and Quantum 
Physics, University of Science and Technology of China, Hefei, 230026, People's 
Republic of China}

\author{Bi-Heng~Liu}
\email{bhliu@ustc.edu.cn}
\affiliation{CAS Key Laboratory of Quantum Information, University of Science 
and Technology of China, Hefei, 230026, People's Republic of China}
\affiliation{Synergetic Innovation Center of Quantum Information and Quantum 
Physics, University of Science and Technology of China, Hefei, 230026, People's 
Republic of China}

\author{Yu~Guo}
\affiliation{CAS Key Laboratory of Quantum Information, University of Science 
and Technology of China, Hefei, 230026, People's Republic of China}
\affiliation{Synergetic Innovation Center of Quantum Information and Quantum 
Physics, University of Science and Technology of China, Hefei, 230026, People's 
Republic of China}

\author{Guo-Yong~Xiang}
\affiliation{CAS Key Laboratory of Quantum Information, University of Science 
and Technology of China, Hefei, 230026, People's Republic of China}
\affiliation{Synergetic Innovation Center of Quantum Information and Quantum 
Physics, University of Science and Technology of China, Hefei, 230026, People's 
Republic of China}

\author{Yun-Feng~Huang}
\affiliation{CAS Key Laboratory of Quantum Information, University of Science 
and Technology of China, Hefei, 230026, People's Republic of China}
\affiliation{Synergetic Innovation Center of Quantum Information and Quantum 
Physics, University of Science and Technology of China, Hefei, 230026, People's 
Republic of China}

\author{Chuan-Feng~Li}
\email{cfli@ustc.edu.cn}
\affiliation{CAS Key Laboratory of Quantum Information, University of Science 
and Technology of China, Hefei, 230026, People's Republic of China}
\affiliation{Synergetic Innovation Center of Quantum Information and Quantum 
Physics, University of Science and Technology of China, Hefei, 230026, People's 
Republic of China}

\author{Guang-Can~Guo}
\affiliation{CAS Key Laboratory of Quantum Information, University of Science 
and Technology of China, Hefei, 230026, People's Republic of China}
\affiliation{Synergetic Innovation Center of Quantum Information and Quantum 
Physics, University of Science and Technology of China, Hefei, 230026, People's 
Republic of China}

\author{Matthias~Kleinmann}
\email{matthias.kleinmann@uni-siegen.de}
\affiliation{Naturwissenschaftlich-Technische Fakult\"at, Universit\"at Siegen, Walter-Flex-Stra{\ss}e 3, D-57068 Siegen, Germany}
\affiliation{Department of Theoretical Physics, University of the Basque 
Country UPV/EHU, P.O.~Box 644, E-48080 Bilbao, Spain}

\author{Tam\'as~V\'ertesi}
\email{tvertesi@atomki.mta.hu}
\affiliation{Institute for Nuclear Research, Hungarian Academy of Sciences, 
H-4001 Debrecen, P.O.~Box 51, Hungary}

\author{Ad\'an~Cabello}
\email{adan@us.es}
\affiliation{Departamento de F\'{\i}sica Aplicada II, Universidad de Sevilla, 
E-41012 Sevilla, Spain}


\begin{abstract}
	We present the first experimental confirmation of the quantum-mechanical prediction of stronger-than-binary correlations. These are correlations that cannot be explained under the assumption that the occurrence of a particular outcome of an $n \ge 3$-outcome measurement is due to a two-step process in which, in the first step, some classical mechanism precludes $n-2$ of the outcomes and, in the second step, a binary measurement generates the outcome. Our experiment uses pairs of photonic qutrits distributed between two laboratories, where randomly chosen three-outcome measurements are performed. We report a violation by {9.3} standard deviations of the optimal inequality for nonsignaling binary correlations. 
\end{abstract}


\date{\today}

\maketitle


{\em Introduction.---}Quantum mechanics is so successful that it is difficult to imagine how to go beyond the present theory without contradicting existing experiments. However, going beyond our present understanding of quantum mechanics can enable us to solve long-standing problems like the formulation of quantum gravity. Some of the most puzzling questions in quantum theory are connected to the measurement process \cite{WZ83}. To go beyond our present understanding of measurements we use recent axiomatizations of quantum theory \cite{Hardy01, DB11, MM11, CDP11, Hardy11} that identify quantum theory as a special case within the general probabilistic theories. We identify an axiom related to the structure of measurements that can be modified in a way not contradicting existing experimental evidence, but making different predictions.


\begin{figure}[t]
\centering
\vspace{-1.8cm}
\includegraphics[width=0.98\linewidth]{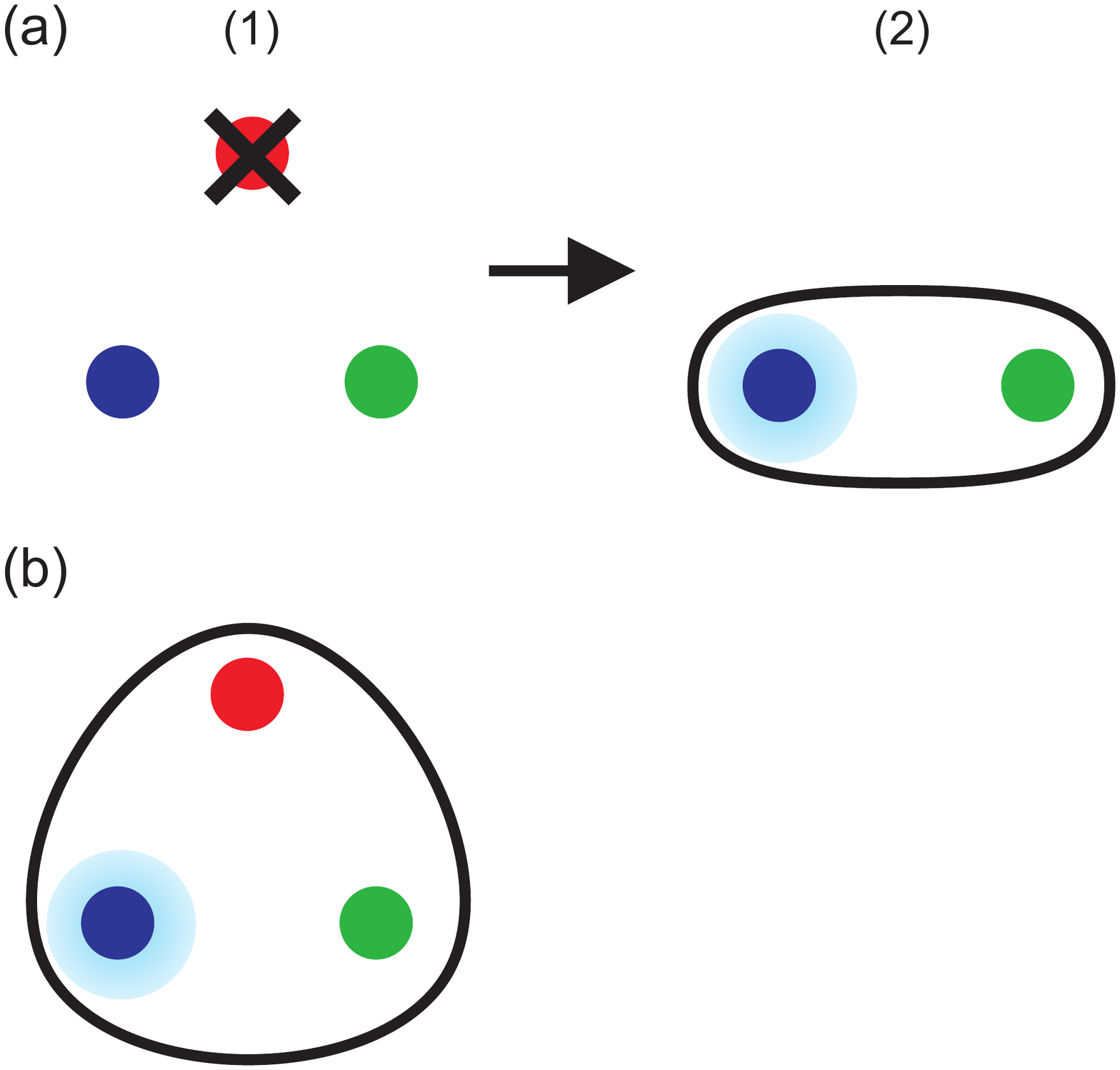}
\vspace{-2cm}
\caption{\label{Fig1}%
Two possible explanations for the measurement process.
Suppose a measurement with three possible outcomes represented by red, green, and blue lights. The process that generates the final outcome (represented by the blue light flashing) can be either (a) a sequence of two steps: (1) The red outcome is precluded by a classical mechanism (e.g., the initial position of the measured system). (2) A general two-outcome measurement selects between the two remaining outcomes. Or (b), the measurement is genuinely ternary in the sense that it cannot be explained as in (a).}
\end{figure}


In quantum theory, two-outcome measurements are described by pairs of operators, $(E, \openone-E)$. A quantum measurement is feasible whenever both operators are positive semidefinite. Conversely, in any general probabilistic theory, if $({\cal E}_1, {\cal E}_2, \ldots, {\cal E}_n)$ represents a feasible $n$-outcome measurement, then any postprocessing to a two-outcome measurement $({\cal E}_1',{\cal E}_2')$ is also a feasible measurement. 
However, according to quantum theory, $(E_1, E_2, \ldots, E_n)$ is already a feasible $n$-outcome measurement whenever all postprocessings to a two-outcome measurement $(E', \openone-E')$ are feasible. This suggests a natural alternative, namely, that feasible $n$-outcome measurements are only those that can be implemented by selecting from two-outcome measurements. Such measurements are hence {\em binary} \cite{KC16} and can be implemented as a two-step process in which, in the first step, some classical mechanism excludes all but two of the outcomes and, in a second step, the final output is produced by a genuine two-outcome measurement. The concept is illustrated in Fig.~\ref{Fig1}.

Correlations between the outcomes of measurements performed by two parties, called Alice and Bob, are described by joint probabilities $P(a,b|x,y)$, where $x$ and $y$ are Alice's and Bob's measurement settings, respectively, and $a$ 
and $b$ are Alice's and Bob's measurement outcomes, respectively. {\em Binary nonsignaling correlations} are those which are both nonsignaling, i.e., $\sum_b P(a,b|x,y)\equiv P_A(a|x)$ and $\sum_a P(a,b|x,y)\equiv P_B(b|y)$, and have only two nontrivial outcomes, i.e., $P_A(a|x)= 0$ except for two outcomes $a$ and $P_B(b|y)= 0$ except for two outcomes $b$, and the convex hull thereof \cite{KC16}. Such correlations also include cases that are forbidden in quantum theory as, for example, Popescu-Rohrlich boxes \cite{PR94} maximally violating the Clauser-Horne-Shimony-Holt inequality \cite{CHSH69}. Interestingly, according to quantum theory, there exist stronger-than-binary nonsignaling correlations \cite{KC16}. A major problem, however, has been to identify how they can be actually observed.

The experiment presented here aims at the maximum violation predicted by quantum theory of the optimal and unique inequality \cite{KVC17} satisfied by binary nonsignaling correlations. The experiment is a bipartite Bell-type experiment in which Alice randomly chooses between two different measurements, $x=0,1$, each of them with three possible outcomes, $a=0,1,2$, and Bob randomly chooses between two different measurements, $y=0,1$, each of them with three possible outcomes, $b=0,1,2$. Binary nonsignaling correlations satisfy the inequality
\begin{equation}\label{ineqa}
 I_a\le 1,
\end{equation}
where
\begin{equation}
 I_a= \sum_{k,x,y=0,1} (-1)^{k+x+y}P(k,k|x,y),
\end{equation}
and the outcomes $a=2$ and $b=2$ do not occur explicitly (see below). In contrast, according to quantum theory, the maximal value 
for $I_a$ is
\begin{equation}\label{qumax}
 I_a= 2(2/3)^{3/2}\approx 1.089.
\end{equation}
This maximum quantum value can be achieved by preparing two qutrits in a particular state and making some particular three-outcome local measurements (see below).

In the experiment we have obtained
\begin{equation}\label{exres}
 I_a=1.066 \pm 0.007
\end{equation}
which implies a violation of the inequality in Eq.~\eqref{ineqa} with a statistical significance corresponding to {9.4} standard deviations. A further analysis of the data (see below) shows that residual systematic errors do not explain this violation.

Consequently, general probabilistic theories in which $n$-outcome measurements are only binary are falsified by showing that there are correlations that are not binary nonsignaling. This also shows that, in nature, there are genuinely 
ternary measurements, thus demonstrating that the measurement process in quantum theory cannot be explained as a two-step process as in Fig.~\ref{Fig1}(a). In fact, the result of the experiment demonstrates that none 
of the four measurements (Alice's or Bob's) can be binary.




{\em Bound on binary nonsignaling correlations.---}The bound $I_a\le 1$ in Eq.~\eqref{ineqa} has been proved in Ref.~\cite{KVC17} by computer-based methods. Here we prove explicitly that the bound $I_a\le 1$ in Eq.~\eqref{ineqa} is valid for binary nonsignaling correlations. We proceed by defining the auxiliary quantities
\begin{subequations}
	\begin{align}
 X_A&= \sum_{a,b,x,y\colon a\ne2} (-1)^{a+x+y} P(a,b|x,y) \text{ and}\\
 X_B&= \sum_{a,b,x,y\colon b\ne2} (-1)^{b+x+y} P(a,b|x,y).
 \end{align}
\end{subequations}
These clearly obey $X_A=0$ and $X_B=0$ for all nonsignaling correlations. We 
then have the inequality
\begin{equation}\label{ineqaux}
 3I_a-X_A-X_B \le\sum_{a,b,x,y}P(a,b|x,y)\equiv 4,
\end{equation}
which holds because the left-hand side of Eq.~\eqref{ineqaux} has only coefficients $\pm1$. Consequently, $I_a\le \tfrac43$ holds for all nonsignaling correlations.

For the bound on binary nonsignaling correlations, it is enough to consider the extremal correlations. By definition, for those there exist certain indices $a_x\in\set{0,1,2}$ for $x=0,1$ and $b_y\in\set{0,1,2}$ for $y=0,1$ such that $P(a,b|x,y)=0$ holds whenever $a=a_x$ or $b=b_y$. The reminder of the entries are then extremal two-outcome correlations and hence are either deterministic, $P(a,b|x,y)\in \set{0,1}$, or they form a Popescu-Rohrlich box \cite{PR94}, implying $P(a,b|x,y)\in \set{0,\tfrac12}$. As a consequence, the bound on $I_a$ must be a multiple of $\tfrac12$ and must not exceed $\tfrac 43$. This proves $I_a\le 1$ for binary nonsignaling correlations. This bound is also tight as can be seen by considering the outcome assignment $a=x$ and $b=2y$.


\begin{figure*}[t]
\centering
\includegraphics[width=\linewidth]{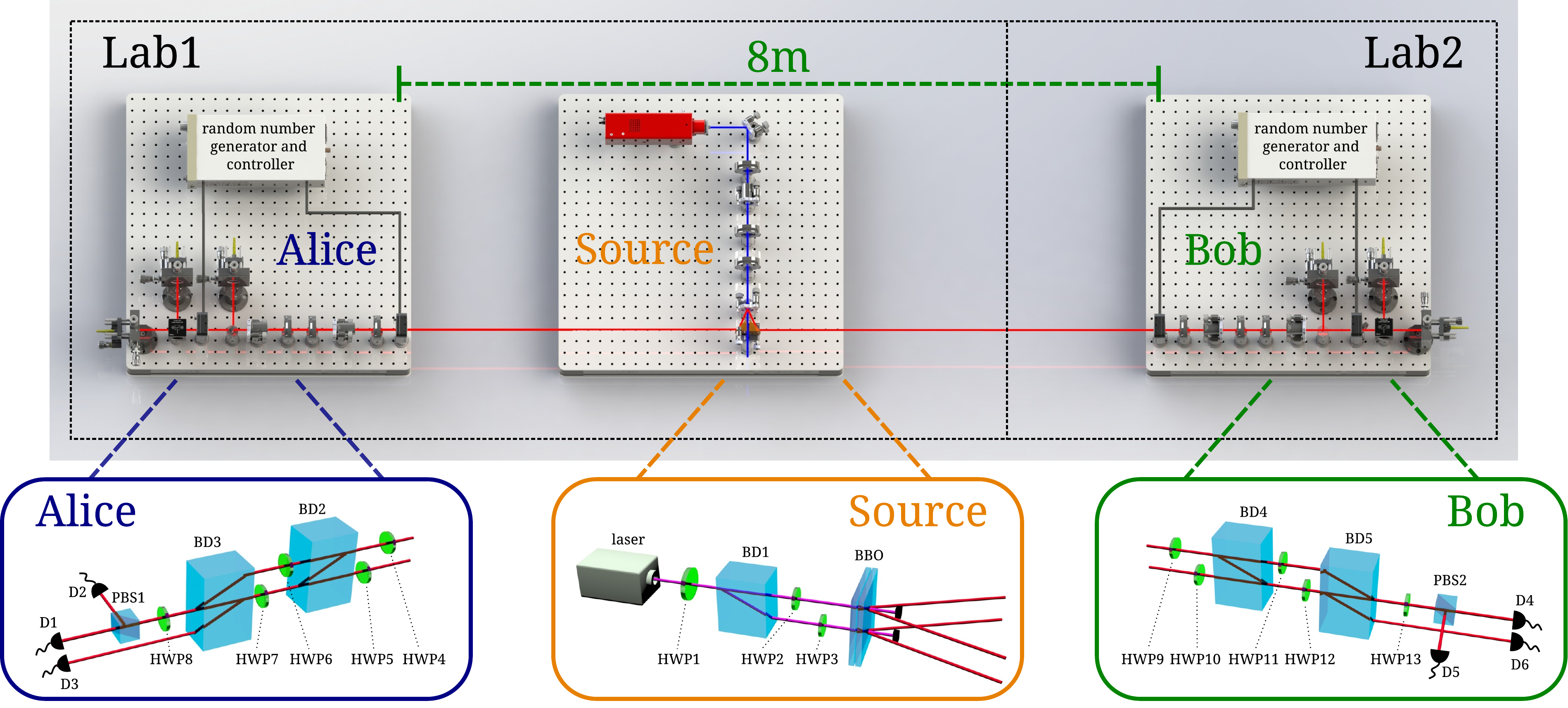}
\caption{\label{Fig2}%
{\bf Experimental setup.}
The source of pairs of photons and the first measurement party, Alice, are in laboratory Lab1, while the second measurement party, Bob, is in laboratory Lab2. The distance between Alice's and Bob's measurement setups is approximately \unit[8]{m}. The pump laser is a continuous wave laser of \unit[404]{nm} wavelength and \unit[100]{mW} power. Subsequently, beam displacers are used to construct phase-stable interferometers. The beam displacers introduce a \unit[4.21]{mm} displacement of the vertically polarized component; beam displacer BD1 operates at \unit[404]{nm} and is approximately \unit[36.41]{mm} long, beam displacers BD2--BD5 operate at \unit[404]{nm} and are approximately \unit[39.70]{mm} long. The pump beam is separated into two paths by means of the half wave plates HWP1--HWP3 and BD1, where the fast axis of HWP1 is oriented at $15\degree$ with respect to the horizontal axis, HWP2 is oriented at $27.37\degree$, and HWP3 at $0\degree$. After BD1 and HWP1--HWP3, the pump state is $(\sqrt{2}\ket{V_u}+ \ket{H_u}- \ket{V_l})/2$, where $H$ ($V$) stands for horizontal (vertical) polarization and $u$ ($l$) denotes the upper (lower) path. The two paths of the pump beam are then focused on two spots of two \unit[0.5]{mm} thick type-I cut $\beta$-borate crystals (BBO) to generate the spatial mode and polarization mode hybrid entangled two-photon state $\ket\psi$; see Eq.~\eqref{state}. The local measurement setting {0}, see Eq.~\eqref{meas0}, and {1}, see Eq.~\eqref{meas1}, for Alice and Bob are constructed using the polarizing beam splitters PBS1 and PBS2, the half wave plates HWP4--HWP13, and BD2--BD4. The orientations of HWP4--HWP13 depend on the measurement setting and are chosen according to Table~\ref{Tab1}. HWP4, HWP8, HWP9, and HWP13 are mounted in electric rotators to switch the measurement settings automatically and the random number generators Quantis-USB-4M (ID Quantique) are used to locally select the measurement basis. Six fiber coupled single photon detectors D1--D6 are used to detect the photons. Interference filters with a bandwidth of $3$ nm are used before each detector to remove background photon noise (not shown). Coincidences between D1--D3 and D4--D6 are detected with the coincidence logic ID800 (ID Quantique, not shown), 
using a coincidence window of \unit[3.2]{ns}.}
\end{figure*}


{\em Experimental test.---}Our experimental setup is described in Fig.~\ref{Fig2} and further develops techniques from Refs.\ \cite{KMWZSS95,HLCZZLLG13,HCLGHZHLG16} optimized for testing the prediction in Eq.~\eqref{qumax}. The source generates the two-photon state
\begin{equation}\label{state}
 \ket\psi= (\sqrt{2}\ket{H_uH_u}+\ket{V_uV_u}-\ket{H_lH_l})/2,
\end{equation}
where $H_u$ ($V_u$) denotes horizontal (vertical) polarization in the upper path and $H_l$ denotes horizontal polarization in the lower path. Consequently, $\ket{H_u}$, $\ket{V_u}$, and $\ket{H_l}$ define a qutrit for Alice and for Bob. The visibility of the entangled state is $0.98\pm0.01$. Each photon of the pair is distributed to a different laboratory and measured there locally.

In each laboratory, the settings {0} and {1} are chosen randomly by means of a random number generator. The measurement outcomes for setting {0} are projectors onto the one-dimensional spaces spanned by
\begin{equation}
\begin{split}\label{meas0}
 \ket{\eta_{0|0}}=&(2\ket{H_u}-(1+\sqrt{3})\ket{V_u}-(1-\sqrt{3})\ket{H_l})
                    /\sqrt{12},\\
 \ket{\eta_{1|0}}=&(2\ket{H_u}-(1-\sqrt{3})\ket{V_u}-(1+\sqrt{3})\ket{H_l})
                    /\sqrt{12},\\
 \ket{\eta_{2|0}}=&(\ket{H_u}+\ket{V_u}+\ket{H_l})/\sqrt3,
\end{split}
\end{equation}
where the projector onto $\ket{\eta_{k|0}}$ corresponds to outcome $k$. 
Similarly, for setting {1},
\begin{equation}
\begin{split}\label{meas1}
 \ket{\eta_{0|1}}=&(2\ket{H_u}+(1+\sqrt{3})\ket{V_u}+(1-\sqrt{3})\ket{H_l})
                    /\sqrt{12},\\
 \ket{\eta_{1|1}}=&(2\ket{H_u}+(1-\sqrt{3})\ket{V_u}+(1+\sqrt{3})\ket{H_l})
                    /\sqrt{12},\\
 \ket{\eta_{2|1}}=&(\ket{H_u}-\ket{V_u}-\ket{H_l})/\sqrt{3}.\\
\end{split}
\end{equation}
These settings together with the state $\ket\psi$ yield the maximal quantum value of $I_a$; see Eq.~\eqref{qumax}. In our setup, the detectors D1--D3 correspond to outcomes $0$--$2$ for Alice and the detectors D4--D6 correspond to outcomes $0$--$2$ for Bob. The measurements are complete with respect to the qutrit space spanned by $\ket{H_u}$, $\ket{V_u}$, and $\ket{H_l}$, while any component of the incoming photon that is outside of the qutrit space remains undetected. In addition, the imperfect efficiency of the detectors, together with the coincidence logic, yield an overall detection efficiency of $0.087\pm0.001$. We account for both effects by implementing the fair sampling assumption, that the coincidences recorded are a representative subsample of what would have been recorded, if all photons were detected.

Data are collected in {4500} runs, with a collection time of \unit[0.5]{s} for each run. Within each run, the measurement settings of Alice and Bob remain fixed. In total, {75$\,$544} coincidences have been recorded.


\begin{center}
	\begin{table*}[t]
		\caption{\label{Tab1}%
			Angles of the fast axis of the half wave plates (HWPs) with respect to the horizontal axis as used in the measurement setups of Alice and Bob; see Fig.~\ref{Fig2}.}
		\centering
		\vspace{1ex}
		\begin{tabular*}{\linewidth}{@{\extracolsep{\fill}}cccccc}
			\hline \hline
			\multirow{2}{*}{Measurement} & HWP4 & HWP5\phantom{0} & HWP6\phantom{0}
			& HWP7\phantom{0} & HWP8\phantom{0} \\
			& HWP9 & HWP10 & HWP11 & HWP12 & HWP13 \\
			\hline 
			Setting 0 (deg)& $-22.5$ & $0$ & $45$ & $17.63$
			& $\phantom{-}37.5$ \\
			Setting 1 (deg)& $\phantom{-}22.5$ & $0$ & $45$ & $17.63$
			& $-37.5$ \\
			\hline \hline
		\end{tabular*}
	\end{table*}
\end{center}


\begin{table}[t]
	\caption{\label{Tab2}%
		$p$-values for (i) joint normalization conditions, (ii) joint nonsignaling conditions, and (iii) the inequality in Eq.~\eqref{ineqN}. ``Coin tosses'' is $q$ if the condition to hold is as plausible as obtaining $q$ times heads in a row when tossing a fair coin. ``Standard deviations'' is $s$ if the condition to hold is as plausible as obtaining a modulus greater than $s$ from a normal distributed random variable.}
\centering
\vspace{1ex}
\begin{tabular*}{\linewidth}{@{\extracolsep{\fill}}lccc}
\hline \hline
 Condition & $p$-value & Coin tosses & Standard deviations\\
\hline
 \multicolumn{4}{l}{Full data set using al 1060 repetitions: $I_a=1.066\pm 0.007$}\\
 (i) & 0.213 & 2.23 & 1.25 \\
 (ii) & 3.66$\times\text{10}^{-\text{4}}$ & 11.4 & 3.56 \\
 (iii) & 5.95$\times\text{10}^{-\text{21}}$ & 67.2 & 9.39 \\
 \multicolumn{4}{l}{Reduced data set using every fifth data set: $I_a=1.08\pm 0.02$}\\
 (i) & 0.340 & 1.56 & 0.954 \\
 (ii)$^a$ & 0.0592 & 4.08 & 1.89 \\
 (iii) & 4.72$\times\text{10}^{-\text{6}}$ & 17.7 & 4.58 \\
\hline \hline
\multicolumn{4}{L{8.8cm}}{$^a$The $\chi^2$ value is unexpectedly below the median of the $\chi^2$ distribution and the $p$-value has been multiplied with a conservative factor of {2}.}
\end{tabular*}\\
\end{table}


{\em Evaluation of the data.---}The {4500} runs with random measurement settings for Alice and Bob, combine to {1060} complete data sets with all four combinations of settings and an average of {67.1} coincidences and for each complete set. We evaluate three conditions on the data:
(i) normalization, i.e., whether $\sum_{a,b} N_r(a,b|x,y)$ is independent of $x$ and $y$;
(ii) nonsignaling, i.e., whether $\sum_a N_r(a,b|x,y)$ is independent of $x$ and $\sum_b N_r(a,b|x,y)$ is independent of $y$;
and (iii) binary correlations, tested by means of the inequality
\begin{equation}
\label{ineqN}\begin{split}
 \sum_{k,x,y=0,1} (-1)^{k+x+y} N_r(k,k|x,y)\\-\frac14\sum_{a,b,x,y}
 N_r(a,b|x,y)\le 0.
\end{split}
\end{equation}
Hereby $N_r(a,b|x,y)$ denotes the number of coincidences for each of the complete data sets $r= 1,\dotsc,1060$ when the outcome of Alice (Bob) is $a$ ($b$) and the setting is $x$ ($y$). We compute the mean $m$ and the variance $v$ over the {1060} runs for each condition, so that $t=m\sqrt{1060/v}$ is distributed according to the Student-$t$ distribution with $g= 1059$ degrees of freedom. In this regime, after rescaling by $\sqrt{(g-2)/g}$, the Student-$t$ distribution is very close to a normal distribution. We therefore obtain the $p$-value of the joint conditions (i) or (ii) using the $\chi^2$ distribution, where there are three independent conditions in (i) and {11} independent conditions in (ii). The obtained values are given in Table~\ref{Tab2} as ``Full data set.''

The full data set shows a violation of the inequality in Eq.~\eqref{ineqN} with a significance corresponding to {9.4} standard deviations. However, also the nonsignaling conditions (ii) are violated by {3.6} standard deviations. The origin of this apparent signaling is the unavoidable fluctuations in the pumping laser. This leads to statistically significant (apparent) signaling since the statistical error is smaller than the error due to the imperfections. A maximum-likelihood fit imposing the nonsignaling constraints increases the value of $I_a$, so that we conclude that the significance of the violation of $I_a$ is nonetheless genuine. To further support this assertion, we reduce the set of samples so that the statistical error is again dominant and consider a reduced data set with only one-fifth of the complete data sets; see Table~\ref{Tab2}, ``Reduced data set.'' There, although the shot noise is increased by a factor of $\sqrt 5\approx 2.2$, a violation of the inequality in Eq.~\eqref{ineqN} by {4.6} standard deviations remains, while the violation of the nonsignaling conditions becomes negligible.

Finally, we compute the empirical frequencies
\begin{equation}
 P_r(a,b|x,y)= \frac{N_r(a,b|x,y)}{\sum_{a',b'} N_r(a',b'|x,y)}
\end{equation}
for each $r$. This allows us to compute for each repetition the value of $I_a$. In Eq.~\eqref{exres} and Table~\ref{Tab2}, we report the resulting mean value and standard error.


{\em Conclusion.---}We have presented an experimental violation with pairs of entangled qutrits of the optimal inequality for nonsignaling binary correlations. Our result (i) provides compelling evidence against two-step explanations of the measurement process, (ii) falsifies nonsignaling binary theories as possible descriptions of nature, apart from the detection and locality loopholes, and (iii) shows, apart from these loopholes, that in nature there exist stronger-than-binary nonsignaling correlations, i.e., correlations that, in particular, cannot be reproduced using Popescu-Rohrlich boxes. The experiment also illustrates how the maturity and refinement achieved by the experimental techniques developed for quantum communication and quantum information processing can be used to test subtle predictions of quantum theory and obtain detailed insights about how nature works.


{\em Data repository.---}The complete data set is publicly available by following the link in Ref.\ \cite{Data}. We encourage readers who want to expand our work with further data analysis to do so.


This work was supported by the National Key Research and Development Program of China (No.\ 2017YFA0304100), the National Natural Science Foundation of China (No.\ 11374288, No.\ 11274289, No.\ 61327901, No.\ 11474268, No.\ 11325419), the Key Research Program of Frontier Sciences, CAS (No.\ QYZDY-SSW-SLH003), the Fundamental Research Funds for the Central Universities, Projects No.\ 
FIS2014-60843-P, ``Advanced Quantum Information,'' and No.\ FIS2015-67161-P, ``Quantum Matter: From Principles to Applications'' (MINECO, Spain), with FEDER funds, the FQXi Large Grant ``The Observer Observed: A Bayesian Route to the Reconstruction of Quantum Theory,'' the DFG (Forschungsstipendium No.\ KL~2726/2-1), the Basque Government (Project No.\ IT986-16), the ERC (Starting Grant No.\ 258647/GEDENTQOPT and Consolidator Grant No.\ 683107/TempoQ), the National Research, Development and Innovation Office NKFIH (Hungary) (No.\ K111734 and No.\ KH125096),
and by the Project ``Photonic Quantum Information'' (Knut and Alice Wallenberg Foundation, Sweden).



\end{document}